\input harvmac
\input epsf

\def\p{\partial}
\def\ap{\alpha'}
\def\half{{1\over 2}}

\def\d{\Lambda}


\def\d3{D_3}
\def\db3{{\bar D}_3}


\Title{}{\vbox{\centerline{Non-Gaussianity in KKLMMT model}}}

\centerline{Qing-Guo Huang$^1$ and Ke Ke$^2$}

\medskip
\centerline{\it $^1$ Interdisciplinary Center of Theoretical
Studies} \centerline{\it Academia Sinica, Beijing 100080, China}
\medskip
\centerline{\it and}
\medskip
\centerline{\it $^2$ Institute of Theoretical Physics}
\centerline{\it Academia Sinica, P. O. Box 2735} \centerline{\it
Beijing 100080}

\bigskip

\centerline{\tt huangqg@itp.ac.cn} \centerline{ \tt kek@itp.ac.cn}

\bigskip

We investigate the non-Gaussianity of the brane inflation which
happens in the same throat in the framework of the generalized
KKLMMT model. When we take the constraints from non-Gaussianity into
account, various consequences are discussed including the bound on
the string coupling, such as the string coupling is larger than
$0.08$ and the effective string scale on the brane is larger than
$1.3 \times 10^{-4} M_p$ in KKLMMT model.

\Date{April, 2005}


\nref\wmapps{H. V. Peiris, et al., Astrophys. J. Suppl. 148 (2003)
213, astro-ph/0302225; D. N. Spergel, et al., Astrophys. J. Suppl.
148 (2003) 175. }

\nref\inf{A. H. Guth, Phys. Rev. D 23 (1981) 347; A. D. Linde, Phys.
Lett. B 108 (1982) 389; A. Albrecht and P. J. Steinhardt, Phys. Rev.
Lett. 48 (1982) 1220. }

\nref\ek{E. Komatsu, et al., Astrophys. J. Suppl. 148 (2003) 119,
astro-ph/0302223. }

\nref\dt{G. R. Dvali and S. H. Tye, Phys. Lett. B 450 (1999) 72,
hep-ph/9812483.}

\nref\brinf{S. Alexander, Phys. Rev. D 65 (2002) 023507,
hep-th/0105032; G. Dvali, Q. Shafi and S. Solganik, hep-th/0105203;
C. Burgess, M. Majumdar, D. Nolte, F. Quevedo, G. Rajesh and R. J.
Zhang, JHEP 07 (2001) 047, hep-th/0105204; G. Shiu and S. H. Tye,
Phys. Lett. B 516 (2001) 421, hep-th/0106274; D. Choudhury, D.
Ghoshal, D. P. Jatkar and S. Panda, hep-th/0305104. }

\nref\fq{F. Quevedo, hep-th/0210292. }

\nref\kklt{S. Kachru, R. Kallosh, A. Linde and S. Trivedi, Phys.
Rev. D 68 (2003) 046005, hep-th/0301240. }

\nref\kklmmt{S. Kachru, R. Kallosh, A. Linde, J. Maldacena, L.
McAllister and S. Trivedi, JCAP 0310 (2003) 013, hep-th/0308055. }

\nref\ft{H. Firouzjahi and S. H. Tye, hep-th/0501099. }

\nref\bkmr{N. Bartolo, E. Komatsu, S. Matarrese and A. Riotto, Phys.
Rept. 402 (2004) 103, astro-ph/0406398. }

\nref\abmr{V. Acquaviva, N. Bartolo, S. Matarrese and A. Riotto,
Nucl. Phys. B 667 (2003) 119, astro-ph/0209156. }

\nref\ks{I. Klebanov and M. Strassler, JHEP 0008 (2000) 052, hep-th/0007191.}

\nref\bkq{C. Burgess, R. Kallosh and F. Quevedo, JHEP 0310 (2003)
056, hep-th/0309187. }

\nref\norm{G. Smoot et al., Astrophys. J. 396, (1992) L1; C. Bennett
et al., Astrophys. J. 464, (1996) L1. }

\nref\ejmmv{K. Enqvist, A. Jokinen, A. Mazumda, T. Multamaki and
A. Vaihkonen, hep-ph/0501076; K. Enqvist, A. Jokinen, A. Mazumda,
T. Multamaki and A. Vaihkonen, hep-th/0502185. }

\nref\mz{J. Minahan and B. Zwiebach, JHEP 0103 (2001) 038,
hep-th/0009246. }

\nref\fgfklt{F. Felder, J. Garcia-Bellido, P. Greene, L. Kofman,
A. Linde and I. Tkachev, Phys. Rev. Lett. 87, 011601 (2001)
hep-th/0012142. }

\nref\stw{G. Shiu, S. H. Tye and I. Wasserman, Phys. Rev. D67:
083517, 2003, hep-th/0207119. }

\nref\cmp{E. Copeland, R. Myers, J. Polchinski, JHEP 0406 (2004)
013, hep-th/0312067; J. Polchinski, hep-th/0412224; B. chen, M. Li
and J. She, hep-th/0504040. }

\nref\cosmics{L. Pogosian, H. Tye, I. Wasserman and M. Wyman, Phys.
Rev. D 68 (2003) 023506, hep-th/0304188; L. Pogosian, M. Wyman and
I. Wasserman, astro-ph/0403268. }

\nref\nong{G. Calcagni, astro-ph/0411773. }

The cosmic microwave background (CMB) data from WMAP \wmapps\
strongly supports that inflation \inf\ should happen before the
hot big bang. The first year results of WMAP also confirm the
emerging standard model of cosmology, a flat $\Lambda$-dominated
universe seeded by nearly scale-invariant adiabatic fluctuations.
By using two statistics, Komatsu et al. \ek\ find that the data
from WMAP consistent with Gaussian primordial fluctuations and
establish limits, $-58<f_{NL}<134$, at $95 \%$ confidence, where
$f_{NL}$ is a non-linear coupling parameter which characterizes
the amplitude of a quadratic term in the primordial potential.

One possible inflation model naturally set up in string theory is
derived by the the potential between the parallel dynamical brane
and anti-brane, namely brane inflation \refs{\dt-\fq}. However the
potential which governs the evolution of inflationary universe
should be flat enough to satisfy the slow roll conditions. In the
usual brane inflation, the tension of the brane (anti-brane) is too
large and the attractive potential between them can be flat enough
only when they separate far from each other. The authors of
\refs{\fq, \kklmmt} pointed out that the distance between the brane
and the anti-brane must be larger than the size of the
extra-dimensional space if the slow roll inflation could happen in
this scenario. This is called $\eta$ problem. Kachru et al. in
\kklt\ successfully introduce some $\db3$ branes in a warped
geometry in type IIB superstring theory to break supersymmetry and
uplift the AdS vacuum to a metastable de Sitter vacuum with lifetime
long enough, but shorter than the recurrence time. If we take an
extra pair of brane and anti-brane in this scenario, a more
realistic slow roll inflation (KKLMMT inflation model) can be
naturally set up \kklmmt. A generalized KKLMMT inflation model has
been also discussed in \ft, where a possible conformal-like coupling
between the scalar curvature and the inflaton is token into account.

In general, $f_{NL}$ contains an order one constant part and a
momentum dependent part, i. e. $g_{NL}({\vec{k_1}},{\vec{k_2}})$
(for example, see \refs{\bkmr, \abmr}). However the present
constraint on non-Gaussianity parameter from WMAP only provides a
constraint on the constant part. Thus the present constraint on
the non-Gaussianity parameter cannot be used to constrain the
inflation at all. However recently the authors of \ejmmv\ found
that the tachyonic instability gives rise to non-Gaussianity
parameter in the primordial perturbation. The tachyon usually
appears in the brane inflation when the inflation ends and the
distance between the brane and anti-brane becomes the same order
as the string length. Therefore the non-Gaussianity can provide a
stringent constraints on the brane inflation model (see the second
paper in \ejmmv).

In this short note, we estimate the non-Gaussianity parameter
$f_{NL}$ and investigate the constraints on the generalized KKLMMT
model in detail. Combining the amplitude of the power spectrum and
the constraints on the non-Gaussianity parameter, we find there is a
stringent constraints on the generalized KKLMMT model.

In KKLMMT inflation model, inflation is drived by the interaction
between a $\d3$ brane and $\db3$ brane which are parallel and
widely separated in five-dimensional AdS space. The $\db3$ brane
is located at the bottom of the throat A with warp factor $h_A$.
The $\d3$ brane is mobile and slowly moves towards the $\db3$
brane due to the attractive force between them. The distance
between the brane and anti-brane plays the role of the inflaton
field. We compactify the string theory on $AdS_5 \times X_5$,
where $X_5$ is a five-dimensional Einstein manifold. The $AdS_5$
solution is given by the metric \eqn\metric{ds^2=h^2(r)
\left(-dt^2+a^2(t)d{\vec x}^2 \right)+h^{-2}(r)dr^2, } with the
warp factor \eqn\wf{h(r)={r \over R}=\exp \left(-{ 2 \pi K \over 3
M g_s} \right), } where $K$ and $M$ are the background NS-NS and
R-R fluxes and $R$ is the curvature radius of the AdS throat \ks,
\eqn\dr{R^4={27 \over 4} \pi g_s N_A \ap^2, } Here $N=KM$ is just
the number of the background $\d3$ charge. Since the curvature
scales like $1/R^2$, $g_s N_A$ should be large in order that the
curvature is small and the supergravity analysis is reliable.

We set an $\db3$ brane at the bottom of throat A with coordinate
$r_0$. Its tension contribute a positive effective cosmological
constant given by \eqn\va{V_A=2T_3 h_A^4,} where \eqn\td{T_3={1
\over (2 \pi)^3 g_s \ap^2}} is the tension of $\d3$ brane. The
force exerted by gravity and the five-form field are of the same
sign and add up, so we have a factor of 2 in eq. \va. The factor
of $h_A^4$ in \va\ is due to a redshift in the curved geometry. Or
from another point of view, we write down the string action in
this background geometry as \eqn\sact{S={M_s^2 \over 2 \pi} \int
d^2 z G_{AB} \p X^A {\bar \p} X^B \sim {(M_s h_A)^2 \over 2 \pi}
\int d^2 z g_{\mu\nu}\p x^\mu {\bar \p} x^\nu, } where
$M_s=1/\sqrt{\ap}$ and $G_{AB}$ is the metric of the target
spacetime, $X^A$ is the coordinate in the target spacetime and
$x^\mu$ is the longitude coordinate on the brane in the metric
\metric. Thus for the observers living on the brane, the effective
string scale should be $M_{obs}=M_s h_A$, which is different from
the string scale in the bulk $M_s$. If the warp factor $h_A$ is
smaller than one, the string seems lighter. This is the key point
why the KKLMMT model can work. In addition, the attractive
interaction also provides a potential \eqn\vdd{V_{\d3 \db3}=-{27
\over 16 \pi^2} {T_3^2 h_A^8 \over \phi^4}, } where
$\phi=\sqrt{T_3}r$ and $r$ is the coordinate corresponding to the
position of the mobile $\d3$ brane. One must ensure that the
compactification volume is stabilized in order to avoid the
decompactification in KKLMMT model. A term coming from the K$\ddot
a$hler potential and various interactions in the superpotential
\kklmmt\ and some possible D-terms \bkq\ should also arise. The
exact formulation of this potential is still not known. Usually it
can be written down as \eqn\vk{V_k = \half \beta H^2 \phi^2, }
which induces a conformal like coupling between the scalar
curvature and the inflaton, here $H$ is the Hubble parameter.
Summing the above potentials up, we obtain the effective potential
in the generalized KKLMMT model, given in \ft\ as follows,
\eqn\pt{V=\half \beta H^2 \phi^2 + 2 h_A^4 T_3 \left(1-{1 \over
N_A} {\phi_A^4 \over \phi^4} \right), } where \eqn\pha{\phi_A^4 =
\left(\sqrt{T_3} r_0 \right)^4={27 \over 32 \pi^2} N_A T_3 h_A^4.
} In eq. \pt, the parameter $\beta$ can be approximately regarded
as a free parameter and the model reduces to KKLMMT model when
$\beta=0$. In \ft, the authors offer the constraints on $\beta$.
Roughly the parameter $\beta$ should satisfy $0 \leq \beta \leq
1/7$ to $1/5$.

This inflation model has been discussed in \ft\ and we directly
quote the useful results. In this model, the slow roll parameter
$\epsilon$ is always small. We need only focus on the slow roll
parameter $\eta$ when we discuss the slow roll conditions. In the
generalized KKLMMT model, $V_A$ dominated the evolution of the
universe. Using eq. \pt, we have \eqn\et{\eta=M_p^2 {V'' \over
V}={\beta \over 3} - {20 \over N_A} {M_p^2 \phi_A^4 \over \phi^6}, }
where $M_p$ is the reduced Planck mass in four-dimensional
spacetime. The inflation ends when the slow roll conditions are
broken down, here $\eta \sim -1$, which gives the final value of the
inflaton field $\phi_{end}$ as \eqn\pend{\phi_{end}^6={20 \over N_A}
{1 \over 1 + \beta/3} M_p^2 \phi_A^4. } The number of e-folds can be
expressed as \eqn\nefd{N_e \simeq {1 \over M^2_p}
\int^{\phi_{N_e}}_{\phi_{f}} {V \over V'} d \phi. } So the value of
$\phi$, namely $\phi_{N_e}$ at the number of e-folding number $N_e$
before the end of inflation is \eqn\pn{\phi_{N_e}^6={24 \over N_A}
N_e M_p^2 \phi_A^4 m(\beta), } where \eqn\mb{m(\beta)={(1+2
\beta)e^{2 \beta N_e} -(1+\beta/3) \over 2 \beta (N_e+5/6)
(1+\beta/3)}. } Now the slow roll parameter can be expressed as
\eqn\srp{\eqalign{\epsilon&={M_p^2 \over 2} \left({V' \over V}
\right)^2={1 \over 18} \left({\phi_{N_e} \over M_p} \right)^2
\left(\beta+{1 \over 2 N_e m(\beta)} \right)^2, \cr \eta&={\beta
\over 3} -{5 \over 6}{1 \over N_e m(\beta)},}} where
\eqn\pmp{{\phi_{N_e} \over M_p}=\left({3^3 \times 5^2 \over 2^5}
\right)^{1\over 4} m^{1 \over 6}(\beta) f^{1 \over 3} (\beta)
N_e^{-{1\over 4}} \delta_H^{\half}, } here we use eq. \pn.

The amplitude of the primordial scalar power spectrum is given by
\eqn\amp{\delta_H={1 \over \sqrt{75}\pi M_p^3}{V^{3\over 2} \over
V'}=\left({2^{11} \over 3\times 5^6 \times \pi^4} \right)^{1/6}
N_e^{5/6} \left({T_3 h_A^4 \over M_p^4} \right)^{1/3} f^{-{2 \over
3}}(\beta), } where \eqn\fb{f(\beta) = \left({2 \beta (N_e+5/6)
\over (1+2\beta)e^{2 \beta N_e}-(1+\beta/3)} \right)^{5/4}
{(1+2\beta)^{3/2} \over (1+\beta/3)^{1/4}} e^{3\beta N_e}. } The
cosmological observations \norm\ show that $\delta_H \sim 1.9 \times
10^{-5}$ at $N_e \sim 55$. In the limit with $\beta \rightarrow 0$,
we find $f(\beta)\rightarrow 1$ and the results are just the same as
the KKLMMT model. Using eq. \amp, we obtain \eqn\thm{g(\beta) \equiv
{T_3 h_A^4 \over M_p^4}={M_{obs}^4 \over M_p^4}{1 \over (2 \pi)^3
g_s}=\left({3\times 5^6 \times \pi^4 \over 2^{11}} \right)^{1/2}
N_e^{-5/2} \delta_H^3 f^2(\beta). } This formula offers a stringent
constraint on the effective string scale on the brane and the string
coupling from the cosmological observations. If we require that the
inflation happens in the throat A, the inflaton should satisfies
\eqn\reginf{\phi_A \leq \phi_{end}<\phi_{N_e} \leq \phi_R, } here
$\phi_R=\sqrt{T_3}R$ and the e-folding number $N_e$ should be
roughly larger than 55 in order to solve the flat and horizon
problem in hot big bang cosmology. Using eq. \pn, \thm\ and \reginf,
we obtain an constraint on the tension of the $\d3$ brane as
\eqn\cdb{{T_3 \over M_p^4} \geq {25 \pi^2 \over N_e } {1 \over N_A}
m^{2 \over 3}(\beta) f^{4\over 3}(\beta)\delta_H^2. } On the other
hand, the reduced Planck scale in the four-dimensional spacetime is
given by \eqn\red{M_p^2 = {2 L^6 \over (2 \pi)^7 \ap^4 g_s^2}, }
where $L$ is the characteristic size of the compactificated space.
The ratio between the $\d3$ brane tension and the Planck energy
density can be expressed as \eqn\rdp{{T_3 \over M_p^4} = {(2
\pi)^{11} \over 4} g_s^3 \left({l_s \over L}\right)^{12}, } here
$l_s=1/M_s$. In order to make the concept of the geometry reliable,
we require that the size of the compactificated space should be
larger than the length of the string, i. e. $l_s \leq L$, which
means that \eqn\tm{{T_3 \over M_p^4} \leq {(2 \pi)^{11} \over 4}
g_s^3. } Combining eq. \cdb\ and \tm, we find a constraint on the
background charge $N_A$ and the string coupling $g_s$ as follows
\eqn\bng{N_A \geq {25 \over (2 \pi)^9} {1 \over N_e} g_s^{-3} m^{2
\over 3}(\beta) f^{4 \over 3}(\beta) \delta_H^2. } This inequality
can be easily satisfied if the string coupling is not too small.
Here we also require that the inflation should end before the
distance between $\d3$ brane and $\db3$ brane reaches string length
$\sqrt{\ap}$, which imposes a new constraints on the string coupling
$g_s$ and the background charge $N_A$ as (see \ft) \eqn\upb{\ln N_A
+ 4 \left( {4 \over 27\pi g_s N_A} \right)^{1/4} \leq {1 \over 3}
\ln \left({2^9 \times 5^2 \times \pi^2 \over 3^3} {1 \over
(1+\beta/3)^2} {1 \over g(\beta)} \right). } Using this inequality,
we can find that there is an up bound for the background charge, e.
g. $N_A \leq 6.9\times 10^6$ for $\beta=0$ and $N_A \leq 1.4\times
10^5$ for $\beta=0.1$. Combining eq. \bng, we can find a low bound
for the string coupling $g_s$, e. g. $g_s \geq 1.2 \times 10^{-8}$
for $\beta=0$ and $g_s \geq 3.6 \times 10^{-6}$ for $\beta=0.1$.

In KKLMMT model, the distance between the $\d3$ brane $\db3$ brane
becomes roughly the same order as the string length and the
tachyon appears when inflation ends. In \ejmmv, the authors point
out that the non-Gaussianity arise due to the instability of the
tachyon and they estimate the non-Gaussianity in the brane
inflation model. Here we estimate the non-Gaussianity in the
generalized KKLMMT model. We propose that the action for the
tachyon in KKLMMT model can be expressed as (the superstring
tachyon for the brane and anti-brane pair, for example, see \mz)
\eqn\sta{S_S=-2T_3 h_A^4\int d^4 x \sqrt{-g}\left({\ap \over 2
h_A^2} e^{-|T|^2/4}\partial_{\mu}T^{*}\partial^\mu
T+e^{-|T|^2/4}\right). } The factor $h_A^4$ appears because of the
the geometry effect in the throat, we should use the effective
string scale instead of the string scale in the bulk. In the
$D\bar{D}$ system, tachyon is a complex filed. Here we may
consider the simplest case where the imaginary part of $T$ is
essentially frozen and only the real part roll down (see \stw\ for
the argument). We redefine the tachyon field as
\eqn\red{\varphi=\int_0^T \sqrt{{2T_3h_A^4 \over M_{obs}^2}}
e^{-T'^2/8} dT'=2\sqrt{{\pi T_3 h_A^4 \over M_{obs}^2}} erf
\left({T \over 2\sqrt{2}} \right)} The transition $T=0\rightarrow
\infty$ corresponds to $\varphi=0 \rightarrow 2\sqrt{{\pi T_3
h_A^4 \over M_{obs}^2}}$. Now the mass square of the tachyon field
$\varphi$ becomes \eqn\tmass{M_\varphi^2=-{M_{obs}^2 \over
2}\left(1-{T^2 \over 4} \right).} When $T>2$, $\varphi$ will not
be a tachyon field any more. So $\varphi$ can be taken as a
tachyon field only when $0 \leq \varphi \leq \varphi_m$ with
$\varphi_m = 2 \sqrt{\pi T_3 h_A^4 \over M_{obs}^2} erf\left({1
\over \sqrt{2}} \right)$. As a result, long wavelength quantum
fluctuations of the tachyon field $\varphi$ with momenta smaller
than $M_{\varphi}$ grow exponentially. As given in \ejmmv, the
fluctuation of the tachyon can be expressed as
\eqn\delphi{\langle\delta\varphi^2\rangle=\int_0^{M_{obs}/\sqrt{2}}{kdk
\over
4\pi^2}e^{2t\sqrt{M_{obs}^2/2-k^2}}={e^{\sqrt{2}M_{obs}t}\left(\sqrt{2}M_{obs}
t-1 \right)+1 \over 16\pi^2t^2}. } The growth of the tachyon
fluctuation continues until $\sqrt{\langle\delta\varphi^2\rangle}$
reaches the value $\varphi_m$, since at $\varphi \sim \varphi_m$
the curvature of the effective potential vanishes and instead of
exponential growth on has the usual oscillations of all the modes.
We can estimate that the time span is $t_*\sim
\left(\sqrt{2}/M_{obs} \right)\ln \left(1/ \sqrt{B g_s}\right)$,
where $B=\left(8 erf^2 \left( 1/\sqrt{2} \right) \right)^{-1}$.
During this period, the number density of the tachyon quanta is
given by \fgfklt \eqn\nk{n_k\sim e^{\sqrt{2}M_{obs}t_*}\sim{1
\over B g_s}. } Thus the number density of the tachyon quanta in
x-space is \eqn\nks{n_\varphi = \int_0^{M_{obs}/\sqrt{2}} {d^3 k
\over (2 \pi)^3}n_k \sim {1 \over (2 \pi)^3 B g_s} \left({M_{obs}
\over \sqrt{2}} \right)^3. }

Just the same as \ejmmv, we assume that the VEV of the inflaton is
vanishing, $\langle \phi \rangle=0$, when the tachyon starts
rolling. The tachyon field and the inflaton field can be divided
into the background, the first order and the second order
perturbation as
\eqn\tap{\eqalign{\varphi&=\varphi_0(\chi)+\delta^{(1)}\varphi(\chi,
\vec{x})+{1 \over 2}\delta^{(2)}\varphi(\chi, \vec{x}), \cr
\phi&=\delta^{(1)} \phi(\chi, {\vec x})+\half \delta^{(2)} \phi
(\chi, {\vec x}), } } where $\chi$ is conformal time. The relevant
part of the metric perturbations is
\eqn\goo{g_{00}=-a(\eta)^2\left(1+2\psi^{(1)}+\psi^{(2)}\right).}
Then the first order and second order perturbation equations can
be written as (see \ejmmv\ for detail form)
\eqn\fpe{{\psi^{(1)}}''-2 A {\psi^{(1)}}'\sim 0, }
\eqn\spe{{\psi^{(2)}}''-2A {\psi^{(2)}}' \sim -{1 \over
M_P^2}\left(2 \left(\delta^{(1)}\varphi' \right)^2 + 8
\varphi_0'^2  \left(\psi^{(1)} \right)^2 -a^2 V_{,\varphi \varphi}
\left(\delta^{(1)} \varphi \right)^2-8\varphi_0' \psi^{(1)}
\delta^{(1)}\varphi'\right), } where $A=\varphi_0''/\varphi_0'$
and the primes denote derivative with respect to the conformal
time $\chi$. We assume that the tachyon modes grow within a time
interval much smaller than the Hubble time and the expansion of
the universe can be neglected, so that we can set $\chi=t$, $a=1$
and $A=\ddot{\varphi_0}/\dot{\varphi_0}$ which can be taken as a
constant. There are two solution with the first order metric
perturbations: $\psi^{(1)}$ is a constant and $\psi^{(1)}\propto
e^{2At}$. We can consider these two cases separately.

If the constant solution dominating, we can take
$\psi^{(1)}\sim10^{-5}$ by using the observed temperature
anisotropies. Now the total energy density stored in $\varphi$ is
\eqn\rhop{\rho_\varphi\sim{1 \over
2}(\delta^{(1)}\dot{\varphi})^2\sim n_\varphi {M_{obs} \over
\sqrt{2}} \sim {M_{obs}^4 \over 32 \pi^3} {1 \over B g_s}.} Now
eq. \spe\ becomes \eqn\sspe{\ddot{\psi^{(2)}} \sim -{2 \over
M_p^2} \left(\delta^{(1)} \dot\varphi \right)^2 \sim -{1 \over 8
\pi^3} {M_{obs}^4 \over M_p^2} {1 \over B g_s},} here we use eq.
\rhop. The second order metric perturbation
\eqn\smp{\psi^{(2)}\sim -{1 \over 8 \pi^3} {1 \over B g_s}
\left({M_{obs} \over M_p} \right)^2 \ln^2 \left( {1 \over \sqrt{B
g_s}} \right). } Thus the standard non-Gaussianity parameter
$f_{NL}$ is given by \eqn\ng{\eqalign{f_{NL}&=-f_{NL}^{\psi}+{11
\over 6}\sim {\psi^{(2)} \over (\psi^{(1)})^2}+{11 \over 6} \cr
&\sim {11 \over 6}+{10^{10} \over 8 \pi^3} {1 \over B g_s}
\left({M_{obs} \over M_p} \right)^2 \ln^2 \left( {1 \over \sqrt{B
g_s}} \right).}} Based on eq. \ng, there is a stringent constraint
on the string coupling $g_s$ and the effective string scale on the
brane $M_{obs}$ from WMAP (see Fig. 1).
\bigskip
{\vbox{{\epsfxsize=9cm
        \nobreak
    \centerline{\epsfbox{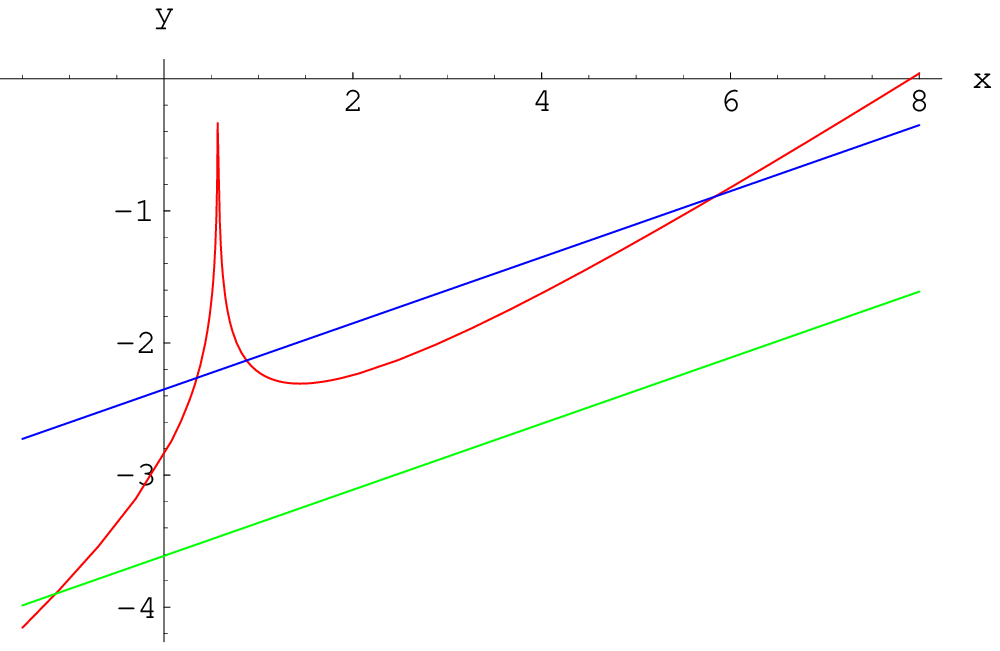}}
        \nobreak\bigskip
    {\raggedright\it \vbox{
{\bf Figure 1.} {\it where $x=\log_{10} g_s$ and $y=\log_{10}
\left( {M_{obs} \over M_p} \right)$. The red line corresponds to
$f_{NL}=134$ and the region below the red line is allowed by WMAP.
The green and blue lines come from the constraints of the
amplitude of the primordial scalar power spectrum (eq. \thm) with
different parameter $\beta$. Here the green line corresponds to
$\beta=0$ and the blue line to $\beta=0.1$. }
 }}}}
    \bigskip}
The region below the red line in Fig. 1 is allowed by the present
constraints from WMAP. Here we set $N_e=55$, $\delta_H=1.9 \times
10^{-5}$, for each $\beta$. Combining the constraints of the
amplitude of the primordial power spectrum \thm, we find that the
non-Gaussianity can give us a stringent constraints on the the
string coupling and the effective string scale on the brane:

1) for $\beta=0$, the string coupling $g_s \geq 10^{-1.1}\simeq
0.08$ and ${M_{obs} \over M_p} \geq 1.3 \times 10^{-4}$. The
constraints on the cosmic F-string and D-string will be:
\foot{Recently it is proposed that cosmic strings can be produced
in KKLMMT model and their effects on cosmological observations are
much studied, see for example \cmp. } $G\mu_F = {1 \over 8 \pi}
\left({M_{obs} \over M_p} \right)^2 \geq 6.7 \times 10^{-10}$ and
$G \mu_D = G T_1 h_A^2 = \left({1 \over 32 \pi g_s} {T_3 h_A^4
\over M_p^4 } \right)^{1/2} \leq 1.3\times 10^{-9}$.

2) for $\beta=0.1$, the string coupling $10^{0.35} \simeq 2.2 \leq
g_s \leq 10^{0.87} \simeq 7.4 $ and $3.0 \times 10^{-4} \leq
{M_{obs} \over M_p}\leq 4.0 \times 10^{-4}$. The constraints on the
cosmic F-string and D-string are: $3.6 \times 10^{-9} \leq G\mu_F
\leq 6.4 \times 10^{-9}$ and $4.6 \times 10^{-8} \leq G\mu_D \leq
8.5 \times 10^{-8}$. OR $g_s \geq 10^{5.8}$ and ${M_{obs} \over M_p}
\geq 6.9 \times 10^{-3}$ and $G \mu_F \geq 2 \times 10^{-6}$,
$G\mu_D \leq 1.6\times 10^{-10}$.

Fitting cosmological constant plus cold dark matter plus strings
to the CMB power spectrum provides an upper limit on the string
tension with $G \mu = {1 \over 8 \pi} (M_{obs}/M_p)^2 < 10^{-6}$
\cosmics. Therefore for $\beta=0.1$, the case with string coupling
larger than $10^{5.8}$ should be ruled out. If we also require
that $g_s \leq 1$ should be allowed, the parameter $\beta$ should
satisfy $\beta \leq 0.03$.

The other extreme case is when the first order metric perturbation
is dominated by the exponential solution $\psi^{(1)} \propto
e^{2At}$. This can happen if the initial fluctuation of the
tachyon field is large enough to overcome the constant solution
for the first order perturbation in the metric. In this case, the
first order perturbation of the tachyon field can be given by
\eqn\phil{\delta^{(1)} \varphi= {2M_P^2 \over
\dot{\varphi}_0}\left(\dot{\psi}^{(1)}+H\psi^{(1)}\right) \sim 4A
{M_P^2 \over \dot{\varphi}_0} e^{2At}. } Substituting this
equation into \spe\ and solving it, we obtain the relevant second
order perturbation as \eqn\ngp{\psi^{(2)}\sim \left(
8+{2M_P^2V^{\prime\prime}(\varphi) \over
\dot{\varphi}_0^2}-{16M_P^2\ddot{\varphi}^2_0 \over
\dot{\varphi}_0^4}-{\dot{\varphi}_0^4 \over M_P^2
\ddot{\varphi}^2_0}\right) e^{4At}. } Thus the non-Gaussianity
parameter can be given by
\eqn\nab{\eqalign{f_{NL}&=-f_{NL}^{\psi}+{11 \over 6} \sim
{\psi^{(2)} \over \left( \psi^{(1)}\right)^2 }+{11 \over 6}\cr
&\sim -{37 \over 6}+16 C g_s \left({M_p \over M_{obs}} \right)^2+
{1 \over C g_s} \left({M_{obs} \over M_p} \right)^2+ 2C g_s
\left({M_p \over M_{obs}} \right)^2\ln^2 \left({1 \over \sqrt{B
g_s}} \right), }} with $C= {2 \pi^2 / erf^2 \left({1 \over
\sqrt{2}} \right) }$. The constraint on the string coupling and
the effective string scale on the brane is shown in Fig. 2.
\bigskip
{\vbox{{\epsfxsize=9cm
        \nobreak
    \centerline{\epsfbox{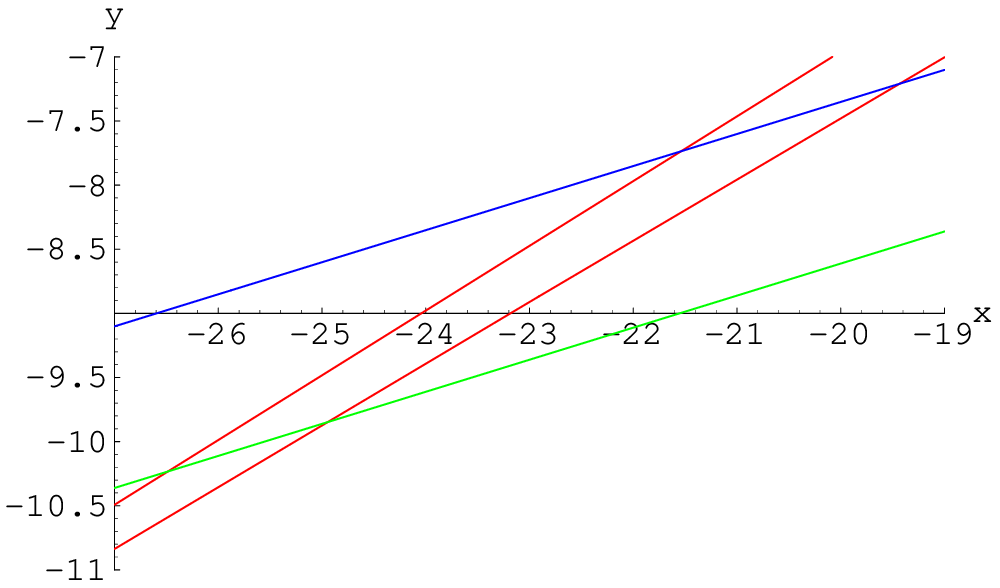}}
        \nobreak\bigskip
    {\raggedright\it \vbox{
{\bf Figure 2.} {\it where $x=\log_{10} g_s$ and $y=\log_{10}
\left( {M_{obs} \over M_p} \right)$. The red line corresponds to
$f_{NL}=134$ and the region between the two red lines is allowed
by WMAP. The green and blue lines come from the constraints of the
amplitude of the primordial scalar power spectrum (eq. \thm) with
different parameter $\beta$. Here the green line corresponds to
$\beta=0$ and the blue line to $\beta=0.1$. }
 }}}}
    \bigskip}
From Fig. 2, the string coupling must be smaller than the limit in
order for the generalized KKLMMT model to work, which we have
obtained before. Thus if the first order metric perturbation is
dominated by the exponentially growing solution, the generalized
KKLMMT model must be ruled by the constraint from the
non-Gaussianity.

In summary, we estimate the non-Gaussianity due to the tachyon
instability in the KKLMMT model. If there is a large first order
metric perturbation due to the fluctuation of tachyon, the
generalized KKLMMT inflation model has been ruled out by the
present non-Gaussianity constraint from WMAP. When the first order
perturbation of metric is not amplified due to the rolling
tachyon, there is still some stringent constraints on the string
coupling and the effective string scale on the brane in the
generalized KKLMMT model. These constraints provide some bounds
for the cosmic F-string and D-string tension. The other
explorations on non-Gaussianity in brane world scenario have also
been discussed wildly (see \nong). We expect that the cosmological
observations in the future, including WMAP, Planck and LIGO, will
offer better opportunity for testing the KKLMMT inflation model.

\bigskip

Acknowledgments

We thank Y.F. Chen, M. Li, J.X. Lu and H. Tye for useful
discussions. The research of Huang is supported by a grant from
NSFC, a grant from China Postdoctoral Science Foundation and and a
grant from K. C. Wang Postdoctoral Foundation.

\listrefs
\end